\documentclass[iop]{emulateapj}

\usepackage{amsmath, amssymb}
\usepackage{hyperref}
\usepackage{url}
\usepackage[]{natbib}
\usepackage{aas_macros}
\usepackage{graphicx}
\usepackage{pagerange}

\usepackage[usenames, dvipsnames]{color}

\newcommand{\appropto}{\mathrel{\vcenter{
  \offinterlineskip\halign{\hfil$##$\cr
    \propto\cr\noalign{\kern2pt}\sim\cr\noalign{\kern-2pt}}}}}

\graphicspath{{figure/}}

\newcommand{\half}{\frac{1}{2}}

\newcommand{\xv}{\bf x}
\newcommand{\yv}{\bf y}
\newcommand{\muv}{\boldsymbol{\mu}}
\newcommand{\pr}{{\rm Pr}}

\slugcomment{LLNL-JRNL-654061}

\shorttitle{Ellipticity Distribution of Blended Objects}
\shortauthors{Dawson, Schneider, Tyson, \& Jee}

\begin{document}

\title{The Ellipticity Distribution of Ambiguously Blended Objects}

\author{William A. Dawson\altaffilmark{1}, Michael D. Schneider\altaffilmark{1,2},  J. Anthony Tyson\altaffilmark{2}, M. James Jee\altaffilmark{2}}

\altaffiltext{1}{Lawrence Livermore National Laboratory, P.O. Box 808 L-210, Livermore, CA, 94551, USA.}
\altaffiltext{2}{University of California, Davis, Physics Department, One Shields Av., Davis, CA 95616, USA}

\email{will@dawsonresearch.com}

\date{Draft \today}



\label{firstpage}

\begin{abstract}
Using overlapping fields with space-based Hubble Space Telescope (HST) 
and ground-based Subaru Telescope imaging we identify a population of blended galaxies that are blended to such a large degree that they are detected as single objects in the ground-based monochromatic imaging, which we label as `ambiguous blends'.
For deep imaging data, such as the depth targeted with the Large Synoptic Survey Telescope (LSST), 
the ambiguous blend population is both large ($\sim14\%$) 
and has a distribution of ellipticities that is different 
from that of unblended objects in a way that will likely be important for the weak lensing measurements.
Most notably, for a limiting magnitude of $i\sim 27$ we find that ambiguous blending results in a $\sim$14\% increase in shear noise (or $\sim12\%$ decrease in the effective projected number density of lensed galaxies; $n_\mathrm{eff}$) due to 
1) larger intrinsic ellipticity dispersion, 2) a scaling with the galaxy number 
density $N_\mathrm{gal}$ that is shallower than $1/\sqrt{N_\mathrm{gal}}$.
For the LSST Gold Sample ($i<25.3$) there is a $\sim7\%$ increase in shear noise (or $\sim7\%$ decrease in $n_\mathrm{eff}$).
More importantly than these increases in the shear noise, we find that the ellipticity distribution of ambiguous blends has an RMS 13\% larger than that of non-blended galaxies.
Given the need of future weak lensing surveys to constrain the ellipticity distribution of galaxies to better than a percent in order to mitigate cosmic shear multiplicative biases, the different ellipticity distribution of ambiguous blends could be a dominant systematic if unaccounted for.
\end{abstract}




\section{Introduction}\label{section:intro}

Object blending becomes progressively worse as the projected surface number density of objects ($n$) in a field increases.
Because future optical surveys, such as the Large Synoptic Survey Telescope\footnote{\url{http://lsst.org}} (LSST) with a limiting r-band magnitude of $\sim27.5$, will be several magnitudes deeper than preceding surveys, we expect that statistical and systematic errors associated with object blending will increase.
\citet{Chang:2013ku} used simulations to confirm that the fraction of blended objects in optical surveys will increase as surveys probe fainter limiting magnitudes and they estimated the increase in shear noise under the assumption that these objects will be rejected from shear analyses.
However, some objects will be blended to such a degree that they will be detected as a single object and cannot be rejected.
We term these \emph{ambiguous blends} and they will be the focus of this paper.

Object blending is a function of the point spread function (PSF), object projected separation ($\theta$), object surface brightness profiles, pixel noise background level, and object number surface density ($n$).
The degree to which objects are blended is a continuum. 
However we find it valuable to define three discrete classes of blends.
We define \emph{ambiguous blends} as two or more blended objects that overlap to such a degree that they are detected as a single object, \emph{conspicuous blends} as two or more blended objects that overlap significantly but are detected as individual objects, and \emph{innocuous blends} as two or more objects that may overlap in the outer periphery of their isophotes but to such a small degree that the overlap has no significant effect on the inferred properties of the objects.

In optical astronomy most existing work has been concerned with the detection and treatment of conspicuous blends \citep[e.g.,][and references therein]{Beard:1990th, Bertin:1996ww}.
For weak lensing, existing studies have been limited to considering how the surface number density of galaxies is decreased after removing conspicuous blends from the sample and the resulting impact on the lensing signal-to-noise ratio (SNR)~\citep[e.g.,][]{Miller:2013bb, Chang:2013ku}.
In other fields of astronomy (e.g., infrared, sub-mm, radio) ambiguous blending has been studied extensively under the topic of \emph{confusion limit}, which is approached when the source density of detected objects per beam (or full-width-half-max) approaches $\sim1/30$ \citep{Scheuer:1957km, Condon:1974kn}.
Most of these studies are concerned with the associated number and flux count uncertainties \citep[e.g.~][]{Oliver:1997we}, although \citet{Hogg:2001cq} considered astrometric uncertainties related to confusion errors.
There have been similar confusion limit studies in the optical that have found that the blue extragalactic background is 29\, mag\,arcsec$^{-2}$ \citep{Tyson:1995}; as the limiting magnitude of surveys approach this limit the effects of ambiguous blending will become larger.
The lack of studies on the effects of ambiguous blending in optical surveys can largely be attributed to the fact that previous surveys were operating at limiting magnitudes much brighter than the optical confusion limit  where other statistical errors (e.g., shape noise) dominated the statistical and systematic errors related to ambiguous blends, which were also smaller due to the relatively small object number densities.

While ambiguous blending related issues are tied to conspicuous blending related issues, 
many aspects of each can be tackled separately.
For example, they will each have different data reduction demands and introduce different systematic errors.
In this article we will focus on ambiguous blending related issues because of the concern that many ambiguous blends may go undetected by standard source extraction and deblending algorithms.
We will consider all other objects `non-blends'.

Ambiguous blending can be expected to act as both an additional source of noise and as well as bias for weak gravitational lensing measurements.
As we will show, ambiguous blending will increase the shear variance both by decreasing the number of objects that can be used to average down the weak lensing shape noise (i.e.~noise due to the intrinsic ellipticity of galaxies), and by contributing more to the shape noise than non-blended galaxies due to the larger scale parameter (i.e.~dispersion) of the ambiguous blend ellipticity distribution.
If the ambiguous blend population has an ellipticity distribution different than than that of the non-blended population (it does as we will show in \S\ref{section:results}), and this distribution remains uncharacterized, then it can lead to a multiplicative shear bias \citep{Hoekstra:2015gh, Bruderer:2015wc}.
This is true for model based shape measurement approaches \citep[e.g.,][]{Bernstein:2013wp, Miller:2013bb} where it take the form of ``model-fitting bias'' \citep{voigt10,bernstein10} due to imperfect knowledge of the unlensed galaxy shapes, as well as approaches that measure the moments of galaxy images \citep[e.g.~the commonly used KSB method][]{Kaiser:1995bi} where the multiplicative biases cannot be correctly calibrated without accurate knowledge of the ellipticity distribution \citep{Viola:2013di}.
\citet{Hoekstra:2015gh} investigated a range of multiplicative biases associated with weak lensing shape measurements as well as their ability to estimate and calibrate these bases and found that, ``the dominant uncertainty in our bias estimate arises from the uncertainty in the ellipticity distribution.''
\citet{Viola:2013di} forecast that the ellipticity distribution dispersion (or scale parameter) will need to be known with a precision of $\sim 0.3$ percent for future weak lensing surveys.

In this paper we use overlapping Subaru SuprimeCam and HST ACS observations to measure the growth in the fraction of ambiguous blends as a function of survey limiting magnitude and show that the ellipticity distribution of ambiguously blended objects (as seen from the ground) is different from that of the remaining objects (\S\ref{sec:method} \& \ref{section:results}), 
show how these results can be understood by considering three fundamental concepts related to blending (\S\ref{section:theory}), 
derive the functional dependence of shear noise estimates on ambiguous blending (\S\ref{sec:noisevsblend}),
and for a given blended population quantify the increase in the shape noise (i.e., shot noise associated with averaging random galaxy shapes with a given ellipticity dispersion) associated with future LSST weak lensing measurements (\S\ref{sec:quantNoiseEffect}).

In what follows it will be important to realize the distinction between different definitions of number density.
From a theoretical standpoint it is most convenient to consider the raw surface number density, $n$, defined at the limiting surface brightness of the survey.
This is similar to what might be observed from a space-based survey, and can be several times larger than the number density observed from the ground, $n_\mathrm{obs}$.
While blending will affect various subsamples differently, it makes most sense to work with $n$ rather than the number density of a particular subsample (e.g.~ the LSST Gold Sample, $i<25.3$) since objects within the subsample will likely be blended with objects outside the subsample (e.g.~$i>25.3$ objects).  

\section{Method of Investigating Ambiguous Blending with Overlapping Ground and Space Based Observations}\label{sec:method}

We utilize fields with overlapping Subaru:SuprimeCam and HST:ACS coverage to test our theoretical expectation that ambiguously blended objects will have an ellipticity distribution with preferentially larger ellipticities than the ellipticity distribution of non-blended galaxies,
and quantify the effect of these blends on the shear noise.
Since the HST PSF is approximately one order of magnitude smaller than the Subaru PSF it enables us to identify ambiguous blends in the Subaru catalog.

\subsection{Data}

We use existing lensing-quality Subaru and HST image data for the Musket Ball Cluster field \citep[see][for data and reduction related details]{Dawson:2012}.
The Subaru data has $0.72\arcsec$ seeing, which is close to the expected mean seeing of future ground based lensing survey telescopes \citep[e.g.~$0.7\arcsec$ for LSST;][]{Collaborations:2009vz}.
After applying quality cuts based on shape error ($0 < \delta e < 0.3$) and magnitude error ($<0.2$), there remain 2356 objects, or $\sim100$ objects per arcmin$^2$, within the approximate HST field of view.
The limiting $i$-band magnitude is $\sim27$.
The HST catalog contains 4637 objects, or $\sim200$ objects per arcmin$^2$, after applying quality cuts based on shape ($\lvert e\rvert<0.9$),  shape error ($0 < \delta e < 0.3$), size (flux radius $>1.2$\,pixels), and magnitude error ($<0.2$).
The limiting F814W magnitude is $\sim28$, and we exclude any objects fainter than 27.
The HST PSF ($0.1\arcsec$) is approximately one order of magnitude smaller than the Subaru seeing enabling us to resolve blended objects.
We summarize the de-blending relevant SExtractor \citep{Bertin:1996ww} parameter specifications used for both datasets in Table \ref{tbl:sextractor}.
We use the ellipticity definition, $\lvert e\rvert\equiv (a-b)/(a+b)$, where $a$ and $b$ are the ellipse major and minor axes, respectively. 
Note that this definition varies slightly from Eqn.~\ref{eq:ellipticity_def}, however the general conclusions of \S\ref{section:ellipdist} will remain unchanged.
Compared with both of the GOODS fields \citep{Giavalisco:2004kl} we estimate that the fraction of ambiguous blends in the Musket Ball Cluster data is $\sim$1.3 times greater than the field.

\begin{deluxetable}{lc}
\tabletypesize{\scriptsize}
\tablecaption{Specified SExtractor de-blending parameters\label{tbl:sextractor}}
\tablewidth{0pt}
\tablehead{
\colhead{Parameter} & \colhead{Specification}
}
\startdata
DEBLEND\_NTHRESH & 8 \\
DEBLEND\_MINCONT & 0.008 \\
FILTER & gauss\_2.0\_5x5.conv \\
CLEAN\_PARAM & 1.2 \\
BACK\_SIZE & 128
\enddata
\end{deluxetable}


\subsection{Blend Identification}\label{sec:blendid}

Our objective for this work is to divide the Subaru objects into blend (i.e.~ambiguous blend) and non-blend (i.e.~remaining) samples.
For each Subaru object we match the closest HST object within a 2$\arcsec$ radius.
These HST objects are defined as \emph{primary matches}.
For each HST object not classed as a primary match, we identify the corresponding Subaru object with the least effective separation,
\begin{equation}
\theta_{\rm eff_{ij}} = \frac{\theta_{ij}}{\Xi_\sigma (\sigma_i + \sigma_j)},
\end{equation}
where $\theta_{ij}$ is the separation of the of HST object $i$ and Subaru object $j$, $\sigma_i$ is the size of the HST object after convolution with a Gaussian kernel 
representative of the Subaru image seeing, $\sigma_j$ is the size of the Subaru object as measured in the Subaru 
image, and $\Xi_\sigma$ is a normalizing scale factor.
When $\theta_{\rm eff_{ij}}<1$ the two objects are separated by less than $\Xi_\sigma$ times the total size of the objects, 
and the Subaru object $j$ is flagged as a potential blend.
Based on visual inspection we find that $\Xi_\sigma = 1$ sets a reasonable criteria for identifying nearly all 
ambiguous blends, with minor non-blend contamination (see \S\ref{section:results} for details).
Finally we visually inspect all Subaru objects flagged as potential blends to increase the purity of our blended sample.
We include side-by-side Subaru and HST images of all visually inspected ambiguous blends in Appendix \ref{sec:ambblendimages}.

\section{Observed Ambiguous Blend Population Properties}\label{section:results}

In this section we use to ambiguous blend catalog from \S\ref{sec:method} to estimate the fraction of ambiguous blends as a function of a survey's limiting magnitude (\S\ref{sec:fracblends}), and compare the ellipticity distribution of the ambiguous blend population with that of the non-blend population (\S\ref{sec:ellipdist}).

\subsection{Fraction of Ambiguous Blends}\label{sec:fracblends}

The automated detection scheme (\S\ref{sec:blendid}) identifies 18\% of the total number of Subaru objects as being part of ambiguous blends.
After visual inspection of these objects in the HST imaging, we confirm that 79\% (341) are actually ambiguous blends, or 14\% of the total Subaru objects.
Most of the false blend detections come from the periphery of the HST field, where there are fewer exposures in the HST images due to dithering, resulting in artifacts such as cosmic rays and noise fluctuations being detected as galaxies.
The next largest class of false blend detections come from objects being improperly segmented during the HST reduction, such as face on spiral galaxies with pockets of bright star formation.

\begin{figure}
	\centerline{
		\includegraphics[width=0.47\textwidth]{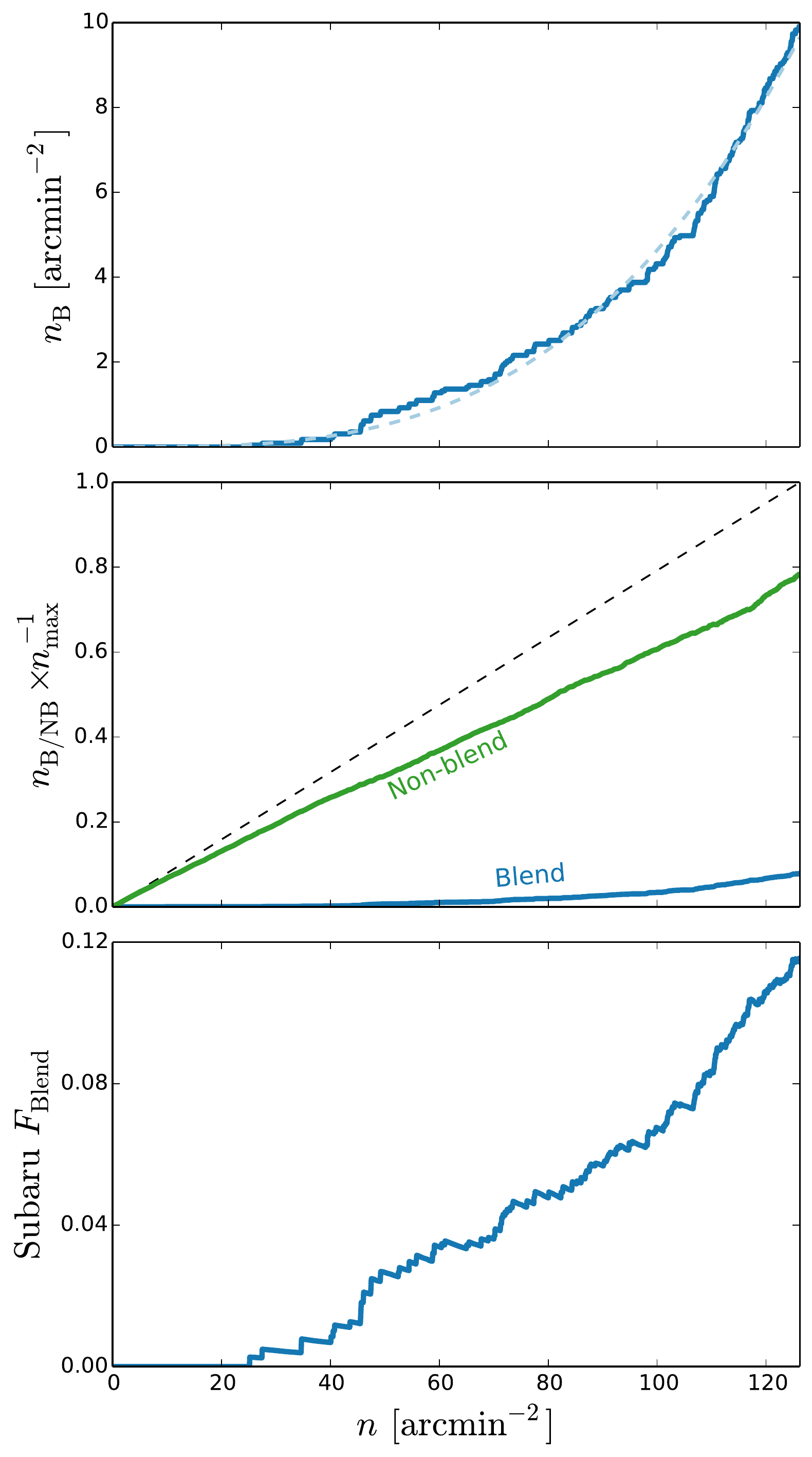}
	}
	\caption{\emph{Top panel}: the cumulative number of ambiguously blended objects per arcmin$^2$ ($n_\mathrm{B}$) in the ground based Subaru observations as a function of the corresponding number density of underlying objects observed by HST ($n$), ordered from brightest to faintest (for F814W $< 27$).
	(Recall from \S\ref{section:intro} that $n$ is different than the number density of a subsample or that observed from the ground.)
	The faint end slope (corresponding to larger $n$) is steeper due to these galaxies being lower signal to noise and thus more likely to contribute to an ambiguous blend, see discussion in \S\ref{section:theory}.
	The light dashed blue line shows the best fit power law 2.3e-6$\times n^{3.15}$.
	\emph{Middle panel}: The number density of ambiguous blends (blue curve) and remaining objects (green curve) relative to the total number density of HST objects. 
	The Blend and Non-blend curves do not sum to give the dashed black  curve since at least two objects are required to make a single ambiguous blend.
	\emph{Bottom panel}: Fraction of Subaru detected objects that are verified ambiguous blends. Compare with Figure 8 of Chang et al.~(2013).
	}
	\label{fig:nblends}
\end{figure}

By examining the ambiguous blends relative to the HST catalog we are able to estimate a number of ambiguous blend properties. 
We find that each Subaru ambiguous blend is on average composed of 2.4 HST detected objects.
This is a heavy tailed distribution though and $\sim75\%$ of the blends are composed of just two objects.
By ordering the HST objects from brightest to faintest we are able to explore how various ambiguous blend and non-blend quantities vary as a function of a survey's limiting raw number density $n$, see Figure \ref{fig:nblends}.
This is done by progressively stepping through the ordered HST catalog and considering correspondingly matched Subaru objects, in this manner a Subaru object transitions from non-blend to blend once two matched HST objects are encountered. 
As can be seen from the dark blue curve in the top panel of Figure \ref{fig:nblends},
we find that the surface number density of ambiguously blended objects  ($n_\mathrm{B}$) increases rapidly as a function of the total raw surface number density ($n$) of observed objects and is well fit by a power law,
\begin{equation}
	n_\mathrm{B} \approx  2.3 n^{3.15} \times 10^{-6},
\end{equation}
which we plot as a dashed light blue curve in the top panel of Figure \ref{fig:nblends}.

\subsection{Ellipticity Distribution of Ambiguous Blends}\label{sec:ellipdist}

\begin{figure}
	\centerline{
		\includegraphics[width=0.47\textwidth]{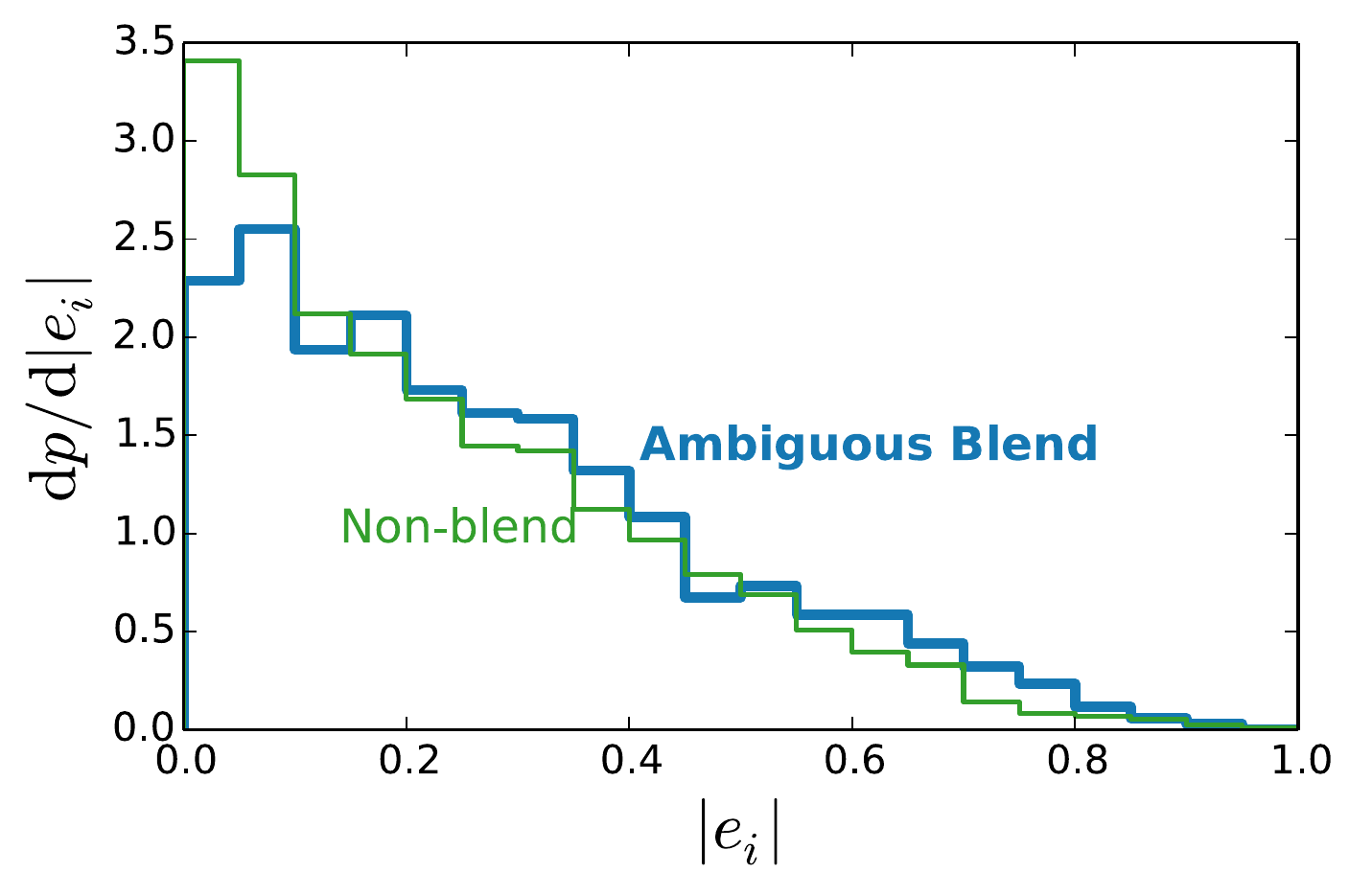}
	}
	\caption{Histograms of the galaxy ellipticity component magnitude as measured from the ground (Subaru). The ambiguous blend ellipticity distribution (thick blue line) is slightly more variable, with $\sigma_\mathrm{SN, B}=0.34$, than the ellipticity distribution of the remainder of the objects (thin green line) with $\sigma_\mathrm{SN, NB}=0.30$.}
	\label{fig:ehist}
\end{figure}

\begin{figure}
	\centerline{
		\includegraphics[width=0.47\textwidth]{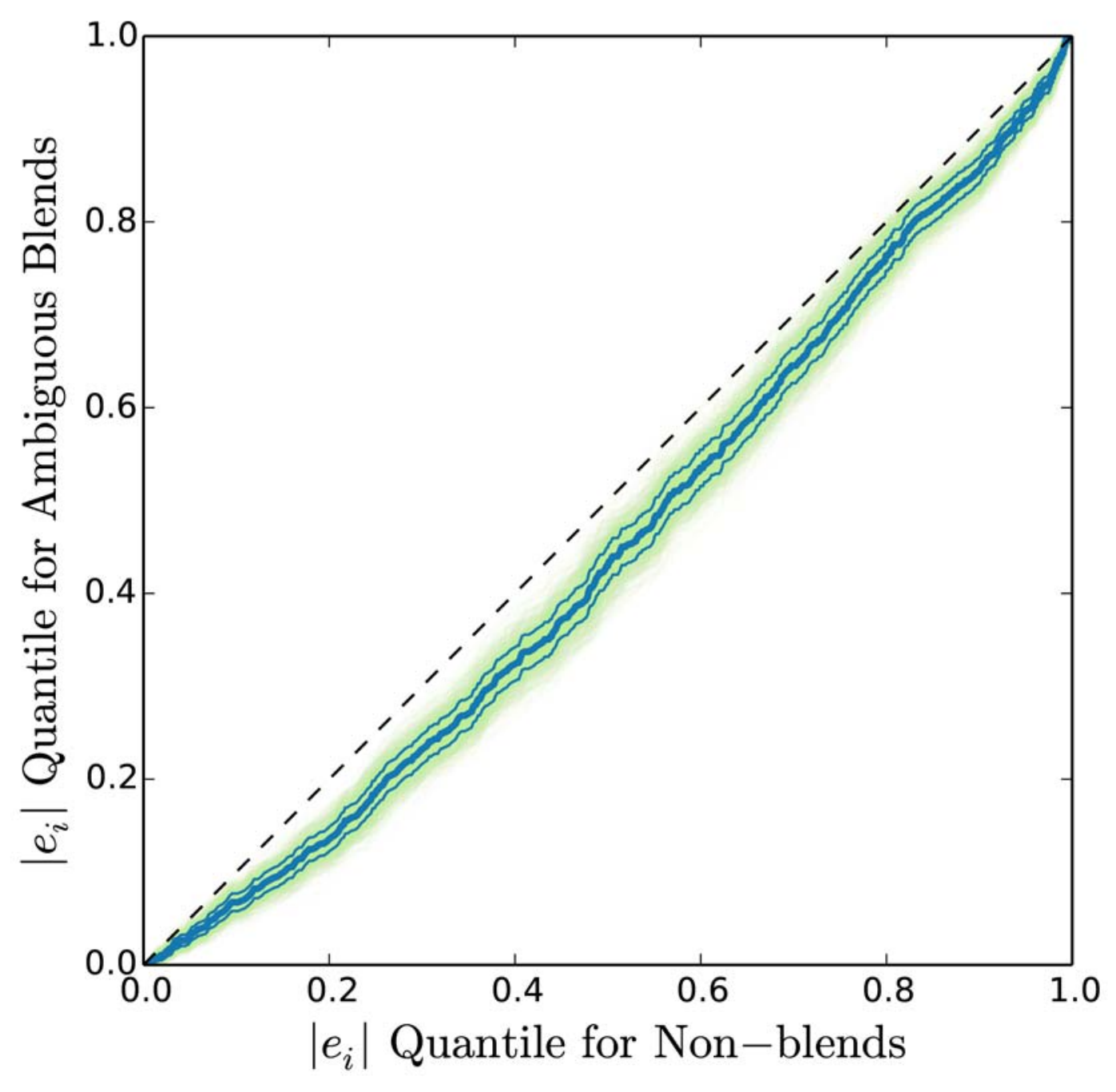}
	}
	\caption{A Q--Q plot (thick blue curve) of the galaxy ellipticity component magnitude $|e_i|$, as measured from the ground (Subaru), for the ambiguous blend population vs.~the remaining population. The thin blue curves show the 68\% deviations based on 1000 bootstrap realizations of the two populations (overlapping green curves). The composite green curves very nearly represent the 3$\sigma$ confidence region. Deviation from the 1:1 dashed line shows that the ambiguous blend population has a significantly different ellipticity distribution than the remaining galaxies.}
	\label{fig:qqplot}
\end{figure}

In Figure \ref{fig:ehist} we compare the ellipticity distributions of ambiguously blended galaxies and non-blended galaxies.
We find that the ambiguous blend populations has a broader distribution of ellipticities with a root mean square (RMS) of 0.34 versus 0.30 for that of the non-blended galaxies.
Since histograms are inexact, in Figure \ref{fig:qqplot} we show a Q-Q plot to quantify the statistically significant difference between the blended and non-blended ellipticity distributions. 
We estimate credible intervals for the quantiles in Figure~\ref{fig:qqplot} 
based on 1000 bootstrap realizations of each population.
We find that the ellipticity distributions of the ambiguous blend and non-blend galaxy populations are different with greater than 5$\sigma$ statistical significance.

\section{Fundamentals of Ambiguous Blending}\label{section:theory}

The findings of \S\ref{section:results} that ambiguous blends will comprise a larger fraction of objects as optical surveys achieve fainter limiting magnitudes and that ambiguous blends will have an ellipticity distribution with larger scale parameter (i.e.~more spread out) than that of non-blended galaxies can be understood by just considering three fundamental relations:
1) pairs of galaxies are distributed such that it is more likely to find galaxies with larger separations (\S\ref{sec:ProjDist}),
2) the probability that two objects will be identified as distinct is zero when their projected separation is zero and remains zero as the separation of the two objects increases up to the point where a deblending threshold is satisfied (\S\ref{sec:BlendDetect}), and
3)  the ellipticity of a pair of objects, which is related to the intensity moments of the two dimensional flux distribution, will increase as the separation (i.e.~moment arm) between those objects increases (\S\ref{section:ellipdist}).
Given relations (1) and (2), ambiguous blends will be composed of objects with preferentially larger separations up to some debelending threshold (\S\ref{sec:ProbAmbBlend}). 
Coupled with relation (3), an ensemble of ambiguously blended pairs of objects will have preferentially larger ellipticities than an ensemble of similar isolated objects.
So despite the complex parameter space in which blending occurs, 
the expectation that ambiguous blends will have an ellipticity distribution with larger scale parameter than that of non-blended galaxies can be understood by just considering the nature of blending as a function of the the two-point projected angular separation $\theta$ and the projected surface density of objects $n$.
We present the supportive details of this rational in the remainder of this section, however readers concerned with the effect of ambiguous blends on weak lensing measurements can continue to \S\ref{section:shearnoisetheory} without loss of continuity.

\subsection{The Projected Distribution of Galaxies}\label{sec:ProjDist}

The probability of finding a pair of galaxies centered in volume elements $dV_1$ and $dV_2$ with separation r is \citep[e.g.,][]{Peebles93},
\begin{equation}
dP_\mathrm{pair, 3D} = n_\mathrm{3D}^2 \left[1 + \xi \left(r / r_0 \right) \right] dV_1 dV_2,
\end{equation}
where the two point correlation function is,
\begin{equation}
\xi = \left( r / r_0 \right) ^{-\gamma},
\end{equation}
$\gamma \approx 1.9$ \citep{Zehavi11}, and $r_0$ is the characteristic clustering length.
This can be expressed in terms of separations along ($r_\parallel$) and perpendicular ($r_\bot$) to the line of sight,
\begin{equation}
dP_\mathrm{pair, 3D} = n_\mathrm{3D}^2 \left[1 + \xi \left(r_\bot, r_\parallel \right) \right] 2\pi r_\bot dr_\bot dr_\parallel.
\end{equation}
However for blending we are interested in the projected distribution of galaxies after integrating along the line of sight dimension
\begin{equation}
dP_\mathrm{pair} \equiv n^2 \left[1 + \omega \left( \theta \right) \right] 2\pi \theta d\theta,
\end{equation}
where $\theta$ is the angular separation of the pair of objects on the sky, and the projected angular correlation function is,
\begin{equation}
\omega = \left( \theta / \theta_0 \right) ^{-\beta},
\end{equation}
$\theta_0$ is a characteristic angular separation, and $\beta = \gamma - 1 \approx 0.9$.
Thus the probability of encountering a pair of galaxies with separation $\theta$ increases approximately linearly with $\theta$,
\begin{equation}\label{eq:prob_pair}
dP_\mathrm{pair} \appropto n^2 2\pi \theta d\theta,
\end{equation}

\subsection{The Ability to Detect Blends as a Function of Projected Separation}\label{sec:BlendDetect}

\begin{figure}
	\centerline{
		\includegraphics[width=0.4\textwidth]{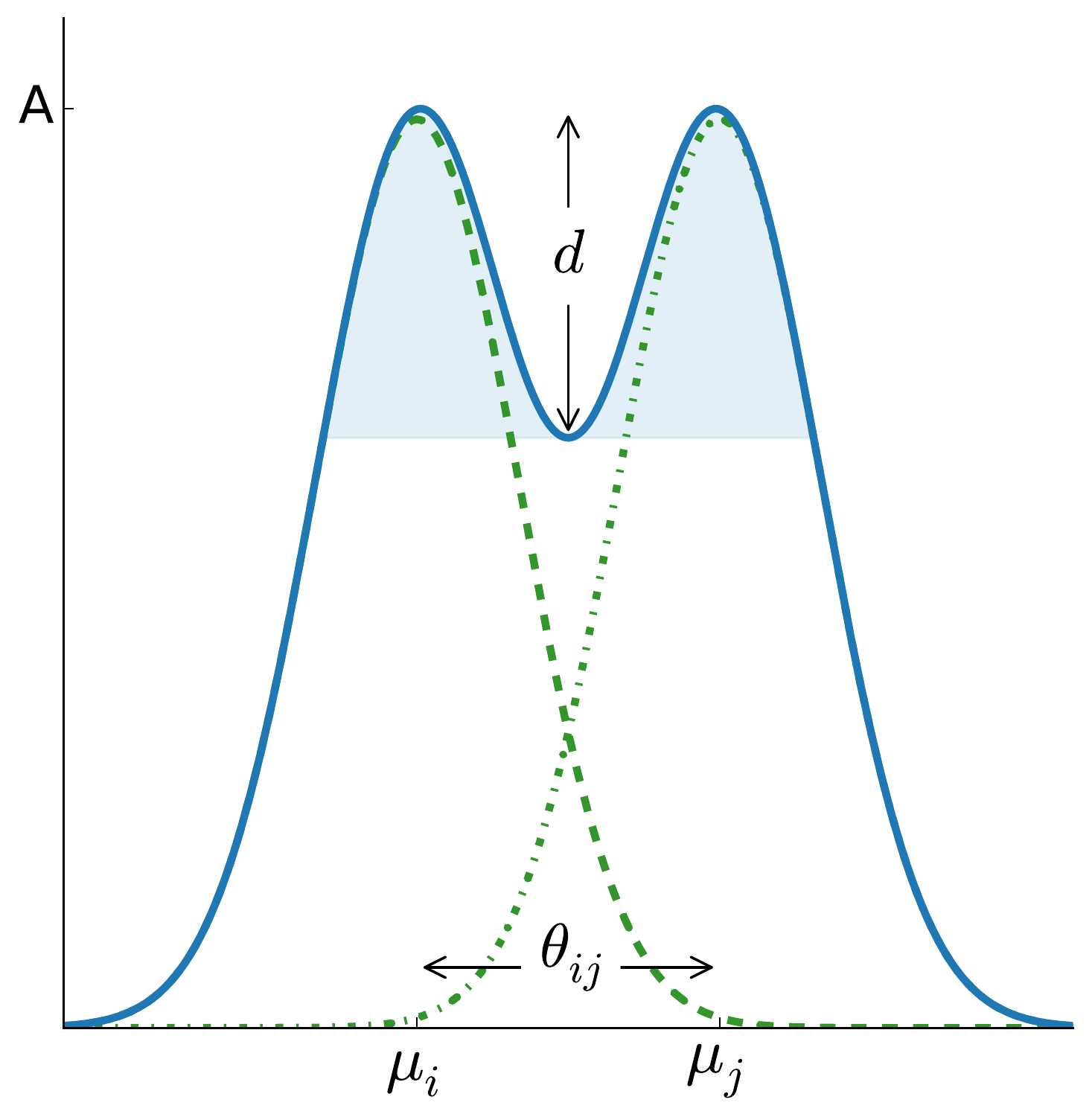}
	}
	\caption{Diagram of two blended galaxies with projected surface brightness profiles (green dash and dot-dash curves), located at $\mu_i$ and $\mu_j$.
	The combined surface brightness profile (blue curve) has a trough depth $d$.
	The shaded blue region represents the integrated intensity of each peak above the saddle point which is compared to the total intensity of the composite object when SExtractor determines whether the composite object is to be considered a blend of two objects (i.e., whether the composite object is an ambiguous or a conspicuous blend).}
	\label{fig:galaxymodel}
\end{figure}

The dominant astronomical source extraction methods for at least the past 38 years all share the basic fundamental approach to blend detection \citep[or deblending; see][]{Beard:1990th, Bertin:1996ww}.
After defining some contiguous region of pixels above a given pixel count threshold,
the local maxima (or peaks) in the region are identified.
Then the decision of whether to consider these peaks as the locations of blended objects is a Boolean function related to the depth of the saddle point between two peaks ($d$ in Figure \ref{fig:galaxymodel}) relative to a given threshold.
Some methods are a direction function of $d$ \citep[e.g.,][]{Jarvis:1981km} and others can be expressed as dependent variables of $d$.
For example, SExtractor \citep{Bertin:1996ww} uses a Boolean function where the integrated pixel intensity of each peak above the saddle point (see shaded area of Figure \ref{fig:galaxymodel}) is required to be greater than a certain fraction of the total intensity of the composite object for it to be considered a blend of two objects.
Regardless of the exact dependence on $d$ all blend identification algorithms share key common traits:
1) When $d=0$ the probability of identifying the two objects as a blend ($dP_\mathrm{detect\,blend}=0$) will be zero,
2) it will remain zero until $d$ increases to the point that a user defined threshold is met and the objects will be identified as a blend ($dP_\mathrm{detect\,blend}=1$).
For all Sersic profiles, $d$ will increase as the projected separation $\theta$ of the two galaxies increases.
Thus the probability of identifying two objects as a blend can be expressed as a Heaviside step function of the projected separation,
\begin{equation}\label{eq:prob_detectblend}
	dP_\mathrm{detect\,blend} = 
	\mathcal{H} \left( \theta-\theta_\mathcal{T}  \right),
\end{equation}
where $\theta$ is the projected separation of the two objects, and $\theta_\mathcal{T}$ is the projected separation when $d$ has increased to the point that the user defined threshold has been crossed and the objects are classified as distinct.

\subsection{Probability of an Ambiguously Blended Pair as a Function of Separation and Projected Density}\label{sec:ProbAmbBlend}

\begin{figure}
	\centerline{
		\includegraphics[width=0.4\textwidth]{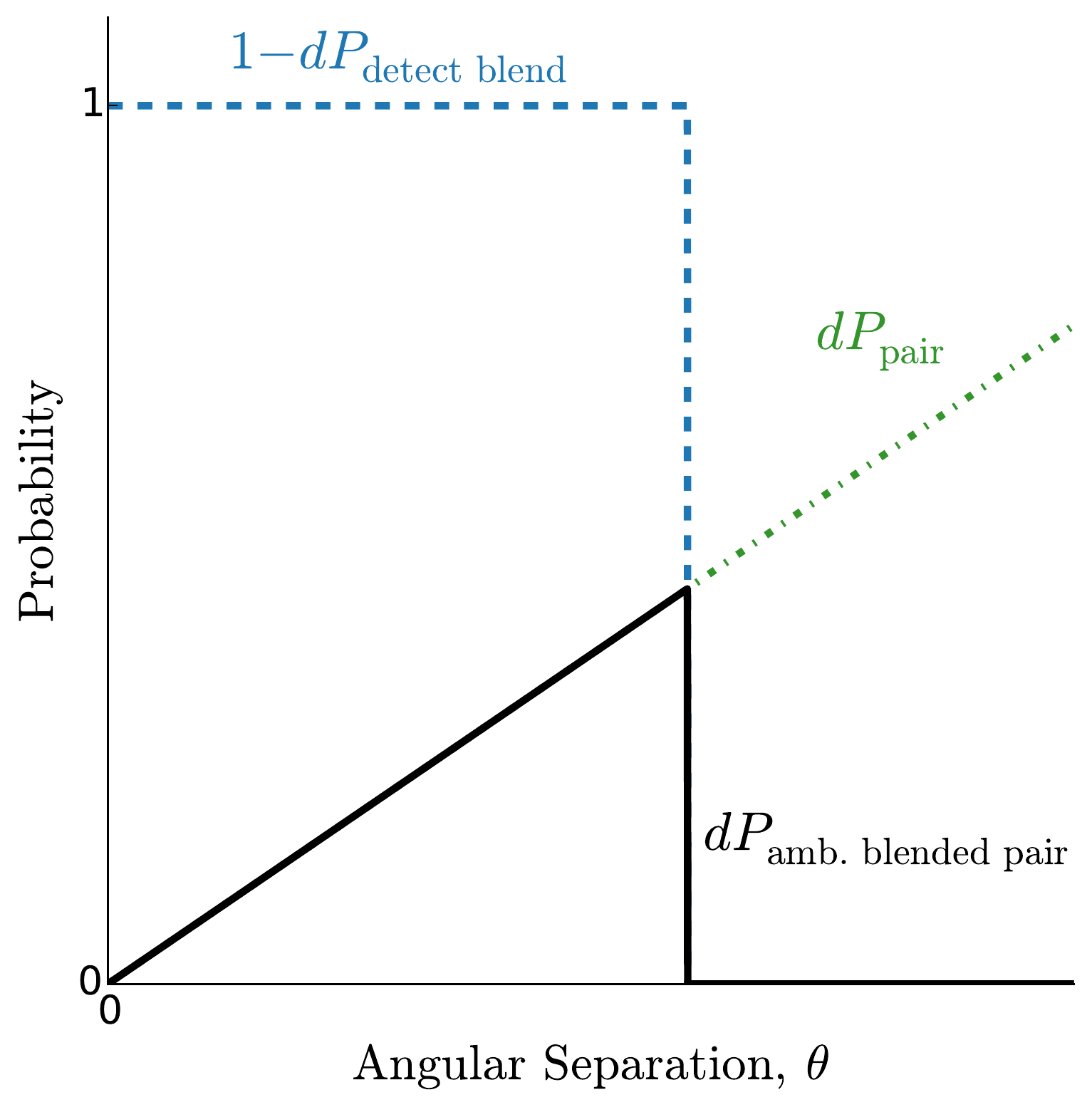}
	}
	\caption{Diagram of the probability distributions involved in determining the ellipticity distribution of ambiguous blends as a function of the galaxy pair angular separation, $\theta$. The probability of there being a pair of galaxies with a given separation (linear green dot-dash curve) joined with the probability of being unable to detect the pair as blended (blue dash curve), results in probability density function for ambiguous blends comprised of galaxies with specified properties throughout the survey. As the ambiguous blend detection threshold $\theta_\mathcal{T}$ is increased (decreased) the location of the step-function will transition to larger (smaller) $\theta$.
}
	\label{fig:probdists}
\end{figure}

Given the probability of encountering a pair of galaxies with projected separation $\theta$ (Equation \ref{eq:prob_pair}) and the ability to disambiguate the pair given that separation (Equation \ref{eq:prob_detectblend}) the probability of observing a pair of ambiguously blended objects is,
\begin{eqnarray}\label{eq:prob_ambblend}
dP_\mathrm{amb.\,blended\,pair} & \propto & dP_\mathrm{pair} \left[1-dP_\mathrm{detect\,blend}\right] \nonumber \\
             & \appropto & n^2 2\pi \theta \left[1-\mathcal{H} \left( \theta-\theta_\mathcal{T}  \right) \right] d\theta.
\end{eqnarray}
The key feature of this probability distribution is that it is zero at $r_{ij}=0$ and increases linearly with slope $n^2$ until a blend detection threshold is reached; see for example the black curve of Figure \ref{fig:probdists}.
Variable galaxy and noise properties will smooth out the peak of the $dP_\mathrm{amb.\,blended\,pair}$ distribution but the general features of a peak offset from zero and $\pr(r_{ij}=0)=0$ will remain.
Similarly these general features are insensitive to the specific deblend threshold. \citet{Jarvis:1982vo} have shown that $\theta_\mathcal{T}$ varies by less than $5\arcsec$ for an ensemble of galaxies over the magnitude range $19\leq J \leq 24$.

\subsection{Ellipticity as a Function of Pair Separation}\label{section:ellipdist}

The standard ellipticity estimator is defined in terms of the pixel intensity quadrupole moments $Q_{kl}$ of the two-dimensional flux distribution $F(\xv)$,
\begin{equation}\label{eq:ellipticity_def}
	\left|e\right|^2 \equiv 
	\frac{\left(Q_{11}-Q_{22}\right)^2+Q_{12}^2}{ \left(Q_{11}+Q_{22}\right)^2},
\end{equation}
where moments are defined as,
\begin{equation}
	Q_{kl} \equiv \int\int dx_1\,dx_2\, x_{k} x_{l}\, F(\xv)
\end{equation}
\cite[see e.g.,][]{Bridle:2009hk}.
Thus as the separation between two galaxies increases (e.g., in dimension 1) the corresponding quadrupole moment in that dimension ($Q_{11}$) will increase, and the composite ellipticity will increase, eventually asymptotically approaching 1 for separations of order the size of the galaxies and larger.
We derive this mathematically in Appendix \ref{sec:eVsSeparation}. The same behavior has been demonstrated with image simulations \citep[see Figure 3 of][]{Jarvis:1982vo}.
As noted in \S\ref{sec:ProbAmbBlend} ambiguously blended galaxies will preferentially have larger projected separations, thus they will have preferentially larger ellipticities compared to isolated non-blended galaxies.

\section{Effect of Ambiguous Blends on Shear Noise}\label{section:shearnoisetheory}

Building on the fundamentals of ambiguous blending (\S\ref{section:theory}) we derive the functional dependence of shear noise on ambiguous blending (\S\ref{sec:noisevsblend}).
Then we use this with the measurements of \S\ref{section:results} to quantify the effect of ambiguous blending on the shear noise (\S\ref{sec:quantNoiseEffect}).

\subsection{Dependence of Shear Noise on Ambiguous Blending}\label{sec:noisevsblend}

In this section we demonstrate why ambiguous blends increase shear noise, and derive an expression that will quantify the increase in shear noise for a given survey number density of galaxies.
In this sub-section we make the following assumptions:
\begin{enumerate}
	\item cosmic shear is estimated as the weighted mean shear $\hat{\gamma}$ of a sample of galaxies,
	\item shape measurement noise ($\sigma^2_{m,i}$; i.e.~the uncertainty associated with ellipticity measurement of any one object $i$) is approximately the same for blended and non-blended galaxies,
	\item the composite objects of an ambiguous blend have been sheared by equal amounts.
\end{enumerate}
With assumption (2), we will disregard shape measurement noise in the remainder of this sub-section.
Assumption (3) disregards potential noise and bias resulting from the blended objects at different redshifts being sheared by different amounts due to the geometric lensing kernel.
While we believe this is a critical issue deserving of future investigation, we consider it important to first elucidate the noise effects of blending for the optimistic scenario where composite source galaxies of an ambiguous blend have been sheared by equal amounts.

With assumption (1), the shear noise can be defined as,
\begin{equation}\label{eq:shear_noise}
	\hat{\sigma}^2_{\gamma}=\frac{\sigma^2_\mathrm{SN}}{n_\mathrm{eff}},
\end{equation}
where $\sigma^2_\mathrm{SN}$ is the intrinsic galaxy shape noise (i.e. the variance of the ellipticity distribution), and $n_\mathrm{eff}$ is the effective number density of galaxies used in the shear measurement \citep[see e.g.,][]{Chang:2013ku, Jee:2014ik},
\begin{equation}
n_\mathrm{eff} = \frac{1}{\Omega}\sum_{i=1}^{N_\mathrm{obs}}\frac{\sigma^2_\mathrm{SN}}{\sigma^2_\mathrm{SN} + \sigma^2_{m,i}},
\end{equation}
where $N_\mathrm{obs}$ is the number of galaxies as observed from the ground, $\Omega$ is the area of the surveyed sky, and $\sigma^2_{m,i}$ is the shape measurement uncertainty of galaxy $i$.
With assumption (2),
\begin{equation}
n_\mathrm{eff} = \frac{N_\mathrm{obs}}{\Omega}\frac{\sigma^2_\mathrm{SN}}{\sigma^2_\mathrm{SN} + \sigma^2_m},
\end{equation}
where $N_\mathrm{obs}/\Omega$ is the projected galaxy number density as observed from the ground ($n_\mathrm{obs}$). 
Then 
\begin{equation}\label{eq:var_withmeasvar}
\hat{\sigma}^2_{\gamma}=\frac{\sigma^2_\mathrm{SN}}{n_\mathrm{obs}} + \frac{\sigma^2_m}{n_\mathrm{obs}}.
\end{equation}
Under assumption (2), the second term of Equation \ref{eq:var_withmeasvar} remains unchanged regardless of the fraction of ambiguous blends.
Furthermore, for most galaxies the shape measurement noise is typically sub-dominant to the shape noise (e.g., in the Subaru:SuprimeCam data that will be discussed in \S\ref{sec:method}, $\tilde{\sigma_m}=0.07$ while $\sigma_\mathrm{SN}\approx0.3$).
In what follows we drop the shape measurement noise term for the sake of simplicity.
We can expand Eqn.~(\ref{eq:var_withmeasvar}) in terms of the separate shape noises of the blended and 
un-blended populations,
\begin{equation}\label{eq:shear_var_def}
	\hat{\sigma}^2_{\gamma} \approx \frac{n_\mathrm{B}\sigma^2_\mathrm{SN,\,B}+n_\mathrm{NB}\sigma^2_\mathrm{SN,\,NB}}{\left(n_\mathrm{B}+n_\mathrm{NB}\right)^2},
\end{equation}
where we now account for the ambiguous blend population (B) and the remaining `non-blend' population (NB) properties, the sum of which account for the number density of observed galaxies $n_\mathrm{obs}$.

Ambiguous blending will cause an increase in the shear noise by: 
1) increasing $\sigma^2_\mathrm{SN}$, due to $\sigma^2_\mathrm{SN,\,B}$ 
being greater than $\sigma^2_\mathrm{SN,\,NB}$, and 
2) decreasing  $n_\mathrm{obs}$ since multiple objects make up a single blend.
To quantify the effects of these contributions we first define $\beta$ to be the 
average number of composite objects per ambiguous blend.
Thus,
\begin{equation}\label{eq:beta_def}
	n_\mathrm{NB}\equiv n-\beta n_\mathrm{B},
\end{equation}
where again $n$ is the raw unblended 
number density of objects for a given limiting magnitude.
Following the rationale of Section \ref{sec:ProbAmbBlend}, the number density of ambiguous blends $n_\mathrm{B}$ is expected to be proportional to some power 
of the total raw number density of observable objects, 
\begin{equation}\label{eq:pl_num_dens}
	n_\mathrm{B}\equiv\nu n^\alpha.
\end{equation}
In the limit that $n$ times the average object area is $\gg1$, then the confusion limit 
has been reached and both $\nu$ and $\alpha\to1$.
In the limit that $n$ times the average object area $\ll1$ 
(as will be the case for most upcoming ground based optical surveys) the number density 
of blends is proportional to the sum of the blending probabilities for each pair of 
galaxies, 
\begin{equation}\label{eq:n_bprop}
	n_\mathrm{B}\propto \int dP_\mathrm{amb.\,blended\,pair},
\end{equation}
where $dP_\mathrm{amb.\,blended\,pair} \propto n^2$ as in \autoref{eq:prob_ambblend}.
Thus, when $n$ is far from the confusion limit,
the index $\alpha$ from \autoref{eq:pl_num_dens} will tend to be $\geq 2$.
It can be greater than 2 in part due to clustering, but to a larger degree it is because $dP_\mathrm{detect~blend}$ is a function of the deblending threshold that is usually defined with respect to the pixel level background.
The number of low SNR galaxies increases as the limiting magnitude increases because 
the number density is observed to scale as 
$n\sim 10^{0.31\left(m-20\right)}$ \citep{Beckwith:2006hp,Tyson:1988km}, 
where $m$ is the limiting apparent $z_{850}$ magnitude of the survey.
Because it is more difficult to detect blends of low SNR objects, 
we should expect larger blend fractions with increasing survey limiting magnitude.
Therefore $\alpha$ should be an increasing function of survey limiting magnitude until it levels off near the confusion limit.

Collecting Equations \ref{eq:shear_var_def}, \ref{eq:beta_def}, and \ref{eq:pl_num_dens}, 
the shear variance including a sub-population of blended galaxies becomes,
\begin{equation}\label{eq:shearvar}
	\hat{\sigma}^2_{\gamma} \approx \frac{\left[\left(1-\beta\nu n^{\alpha-1}\right)\sigma^2_\mathrm{SN,\,NB}+\nu n^{\alpha-1}\sigma^2_\mathrm{SN,\,B}\right]}{n\left[1-\nu n^{\alpha-1}\left(\beta-1\right)\right]^2}.
\end{equation}
Both $\beta$ and $\sigma^2_{\rm SN, B}$ are functions of the threshold defining 
when objects are classified as distinct or blendend. 
In \S\ref{section:results} we fit for the values of $\beta$, $\nu$, and $\alpha$, showing the 
blend fraction $n_\mathrm{B}$ is well-described by the power-law model in \autoref{eq:pl_num_dens}.
\autoref{eq:shearvar} then provides a new model for the shear variance that can be used in cosmic shear 
analyses of surveys with non-negligible blending.

\subsection{Quantifying the Effect of Ambiguous Blending on Shear Noise}\label{sec:quantNoiseEffect}

By using the observed fraction of ambiguous blends in the Subaru image relative to the HST image (\S\ref{sec:fracblends}) we are able to estimate the $\alpha$, $\beta$, and $\nu$ parameters discussed in \S\ref{section:theory}.
We find that $\beta\approx2.4$, meaning each Subaru ambiguous blend is on average composed of 2.4 HST detected objects.
Fitting a power law to the $n_\mathrm{B}$ curve of Figure \ref{fig:nblends} we find that 
$\nu$=(2.3$\pm$0.07)$\times10^{-6}$ and $\alpha=3.15\pm0.007$, with $\Sigma_{\nu\alpha}=-4.9\times10^{-10}$.

Referring to Equation \ref{eq:shearvar}, and given the $\alpha$, $\beta$, and $\nu$ parameter estimates as well as 
the RMS ellipticity of the blended and non-blended populations (\S\ref{sec:ellipdist}) we find that blending results in a shear noise increase of $\sim$14\%.
This is equivalent to a 12\% decrease in the effective number density of galaxies $n_\mathrm{eff}$. 
Recall from \S\ref{section:shearnoisetheory} that a number of optimistic assumptions have been made in this estimate so it should be considered a lower limit on the expected shear noise increase due to ambiguous blending.

We now consider the sample of Subaru galaxies which would qualify as the LSST Gold Sample ($i<25.3$), since these are what will eventually be used for the primary cosmic shear measurements \citep{Collaborations:2009vz,Collaboration:2012uk}.
While we only use galaxies with $i<25.3$ to estimate the ambiguous blend and non-blend population properties, we consider all galaxies ($i\lesssim27$, which is approximately the LSST limiting magnitude) when determining which of the $i<25.3$ galaxies are ambiguous blends.
We still find that the blend and non-blend ellipticity distributions are significantly different in a Q-Q plot, however we find that the RMS ellipticity of the blended populations reduces from 0.34 to 0.32 for the Gold Sample compared to the full depth sample.
Below $|e_i|\sim0.5$ the blend distribution looks nearly identical to that in Figure \ref{fig:ehist}, however above $|e_i|\sim0.5$ there is almost no difference between the blend and non-blend population.
For this subsample we find that ambiguous blending results in a shear noise increase of $\sim7\%$ versus a sample 
without ambiguous blends, which is equivalent to a 7\% decrease in the effective number density of galaxies.
This suggests that, upon further investigation, the blending related shear noise metric in Equation \ref{eq:shearvar}
could be coupled with other lensing quality metrics to determine an optimal LSST Gold Sample.

\section{Summary \& Conclusions}

We have shown that, despite the complex parameter space of object blending, the expectation that ambiguously blended objects will have a significantly different ellipticity distribution from that of non-blended objects can be understood simply by considering the functional dependence of ambiguous blends on the projected separation of objects (\S\ref{section:theory}).
Using similar rationale we have shown that the number density of ambiguous blends is expected to be proportional to some power of the total raw projected number density of objects (Equation \ref{eq:pl_num_dens}) with power $\geq 2$.
We have also formulated how the variance of standard shear estimators will be affected by ambiguous blends (Equation \ref{eq:shearvar}) and have shown that it will increase both due to the reduction of the observed number density as well as the increase in the shape noise (i.e., from larger ellipticity distribution variance) relative to the non-blended population.

We use overlapping Subaru SuprimeCam and HST ACS observations to empirically confirm these expectations and quantify the parameters of the ambiguous blend formalism ($\alpha,~\beta,~\nu$; see \S\ref{section:shearnoisetheory}) for upcoming ground based surveys such as LSST.
We find that 14\% of the objects in the ground based imaging are ambiguous blends, and that these objects have a larger RMS ellipticity distribution compared to the non-blended objects, 0.34 vs.~0.3.
We find that the number density of ambiguously blended objects increases rapidly as a function of the total raw number density of observed objects,
\begin{equation}
	n_\mathrm{B} \approx  2.3 n^{3.15} \times 10^{-6}.
\end{equation}
We also find that, on average, ambiguous blends are composed of 2.4 objects (i.e.~$\beta=2.4$). However, this is a heavy-tailed distribution and $\sim75\%$ of the ambiguous blends are composed of just two objects.
So it seems an appropriate approximation for future ambiguous blending studies to first focus on two-object blends.

Given our parameter estimates we find that for a survey with a limiting magnitude of $i\sim 27$, ambiguous blending will result in a $\sim14\%$ increase in shear noise (or $\sim12\%$ decrease in $n_\mathrm{eff}$).
If just a sub-sample of these galaxies are considered ($i<25.3$, representative of the LSST Gold Sample) then there is a $\sim7\%$ increase in shear noise (or $7\%$ decrease in $n_\mathrm{eff}$).
Thus, even when we optimistically ignore the systematic biases associated with ambiguous blending, it will have a non-negligible impact on the power of future weak lensing surveys.  

While the increase in shear noise due to ambiguous blends is non-negligible, our findings indicate that a more serious problem is the multiplicative shear bias associated with ambiguous blends.
A we discussed in \S\ref{section:intro}, uncertainty in the ellipticity distribution of lensed galaxies can be the largest source of multiplicative bias in weak lensing measurements \citep{Hoekstra:2015gh}, and it is forecast that the ellipticity distribution will need to be known with a precision of $\sim3\%$ for future weak lensing surveys \citep{Viola:2013di}.
Given our finding that ambiguous blends have a significantly different and $13\%$ larger RMS ellipticity distribution compared to non-blended objects, suggests that ambiguous blends could result in a limiting systematic bias for future ground based weak lensing surveys if improperly accounted for.
For example, if an ellipticity distribution were used that failed to account for ambiguous blends and assumed that all observed objects were non-blends (as might be the case if the ellipticity distribution were measured from space-based observations), then the simulations of \citet{Hoekstra:2015gh} combined with our results show that the residual multiplicative bias after calibration could be $\sim0.02$, which is an order of magnitude larger than the requirements of future surveys \citep{Mandelbaum:2014vw}.

The majority of existing studies related to blending have focused on deblending conspicuous blends \citep[e.g.,][and references therein]{Beard:1990th, Bertin:1996ww} or assessing the impact of excluding those objects from the analyses \citep[e.g.,][]{Chang:2013ku}.
However, our findings suggest that the little studied subject of ambiguous blending is a potentially more important consideration for future surveys than conspicuous blends.
In the interest of advancing future studies of ambiguous blending we highlight some of the limitations of this study:
\begin{itemize}
\item We have only directly considered the impact of ambiguous blending on the weak gravitational lensing shear variance.
As we noted in \S\ref{section:intro}, ambiguous blending will also introduce a number of biases for gravitational shear measurements, as well as impact a wide range of other astrophysical and cosmological measurements.
\item Our study surveys a field containing a galaxy cluster. We estimate that the fraction of ambiguous blends in this field is $\sim 1.3$ times greater than the average over randomly chosen areas of the sky.
\item We have only considered a single empirical $0.72\arcsec$ seeing. Both the fraction and ellipticity distribution of ambiguous blends will vary as a function of seeing. It is worth considering the possibility of using optimal seeing ($\sim 0.5\arcsec$) images from a multi-epoch survey to identify some of the ambiguous blends in the full multi-epoch stack.
\item We have only considered a monochromatic detection band. It is likely that a multi-band photometric detection scheme will help identify a sub-population of ambiguous blends which are composed of different color objects.
\item We have not considered how the effective shear responsivity \citep[see e.g.,][]{Bridle:2009hk} distribution of ambiguously blended objects will differ from that of non-blended objects.
\item We have used SExtractor \citep{Bertin:1996ww} to identify objects. A source extraction algorithm trained to the possibility of ambiguous blends could flag some of them.
\end{itemize}


\section*{Acknowledgments}
We thank the LSST DESC members for many valuable conversations related to this work, in particular David Kirkby and Andrew Bradshaw. 
Part of this work performed under the auspices of the U.S. DOE by LLNL under Contract DE-AC52-07NA27344.
This material is based upon work supported by the NSF under Grant No. AST-1108893 and DOE under grant DE-SC0009999.
Support for program number GO-12377 was provided by NASA through a grant from STScI, which is operated by AURA, under NASA contract NAS5-26555.
Based in part on data collected at Subaru Telescope, which is operated by NOAJ.


\appendix

\section{Images of Ambiguous Blends}\label{sec:ambblendimages}
\newcommand{\noprint}[1]{}
\newcommand{\figsetstart}{{\bf Fig. Set} }
\newcommand{\figsetend}{}
\newcommand{\figsetgrpstart}{}
\newcommand{\figsetgrpend}{}
\newcommand{\figsetnum}[1]{{\bf #1.}}
\newcommand{\figsettitle}[1]{ {\bf #1} }
\newcommand{\figsetgrpnum}[1]{\noprint{#1}}
\newcommand{\figsetgrptitle}[1]{\noprint{#1}}
\newcommand{\figsetplot}[1]{\noprint{#1}}
\newcommand{\figsetgrpnote}[1]{\noprint{#1}}
 
\figsetstart
\figsetnum{6}
\figsettitle{Ambiguously Blended Images}

\figsetend

\begin{figure}
\figurenum{6}
\centerline{
	\includegraphics[width=\textwidth]{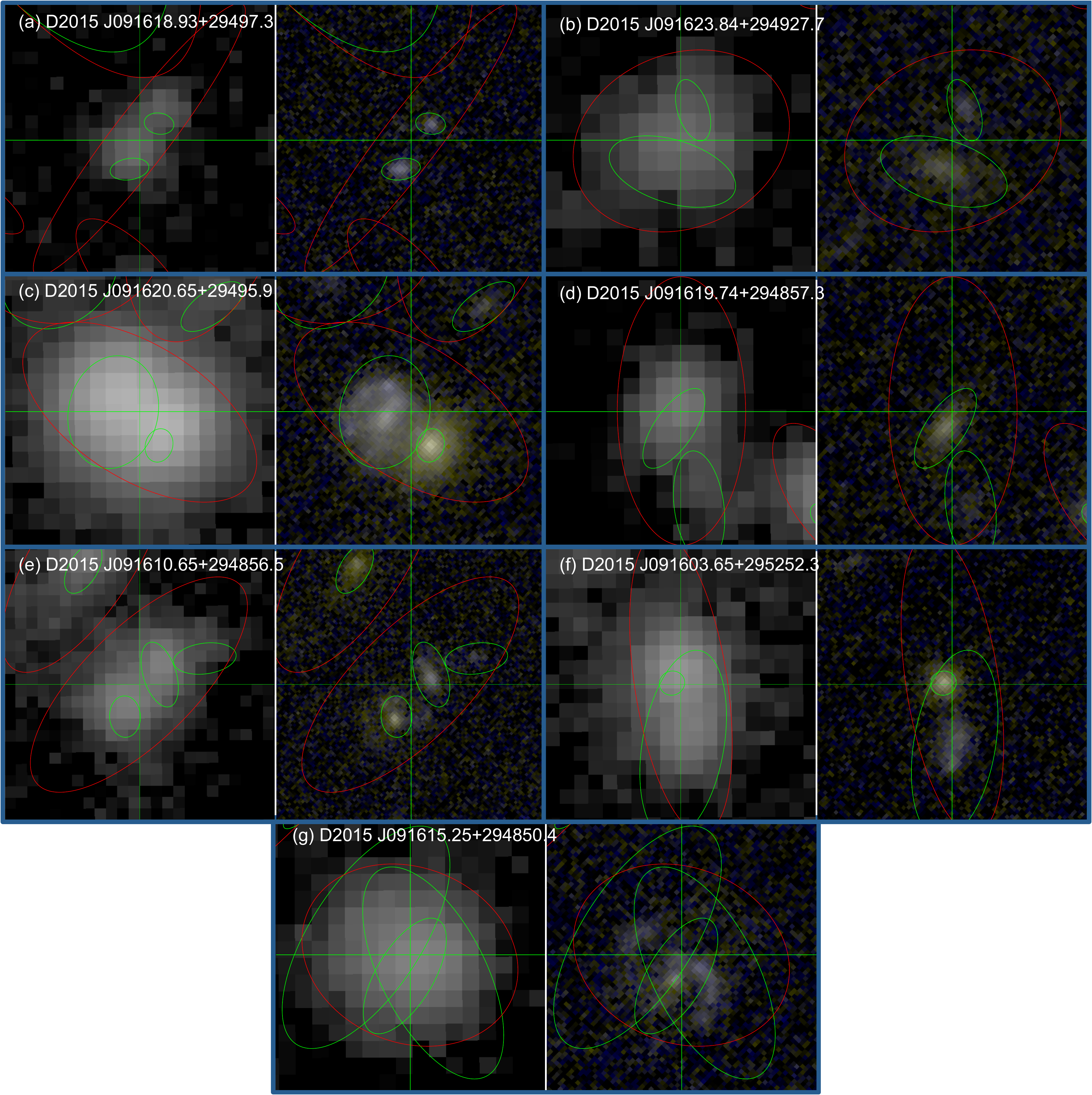}
}
\caption{
Visually confirmed ambiguous blends in the Musket Ball Cluster Subaru/HST field (Dawson et al. 2013). For each blend, the Subaru i-band image (left) is shown alongside the HST color image (right; b=F606W, g=F814W, r=F814W). Both images are logarithmically scaled. The ellipses show the observed object ellipticities (red = Subaru, green = HST). The images and green crosshair are centered on the Subaru ambiguous blend object center. The Subaru pixel scale is 0.2 arcsec/pixel, and the HST pixel scale is 0.05 arcsec/pixel.
Panels (a)-(g)  show blends selected from the complete sample (available in the electronic edition of the article) to highlight some of the common ``classes'' of ambiguous blends.
Panel (a) is an example of a case where two objects with small ellipticity have become ambiguously blended in the Subaru image and produced a single detected object with large ellipticity (Subaru object FWHM: $1.6\arcsec$).
Panel (b), while less common, it is also possible to have two objects be ambiguously blended together to create a smaller ellipticity object observed in Subaru (Subaru object FWHM: $1.0\arcsec$).
Panel (c) is an example of two objects with similar brightness that are ambiguously blended (Subaru object FWHM: $1.3\arcsec$).
Panel (d), two objects need not have similar brightness to generate an ambiguous blend with significantly different ellipticity properties compared to that of the brighter object. Even objects in the LSST Gold Sample ($i < 25.3$) can be significantly affected by the fainter objects ($25.3 < i < 28$) in the survey (Subaru object FWHM: $1.2\arcsec$).
Panel (e), approximately 25\% of ambiguous blends are composed of more than two objects (Subaru object FWHM: $1.8\arcsec$). 
Panel (f), is an example of two objects, likely at different redshifts (given their different colors and magnitude), that are ambiguously blended (Subaru object FWHM: $1.4\arcsec$).
Panel (g), may be a spiral galaxy that has become fragmented during the reduction of the HST imaging, thus it may be an example an artificial ambiguous blend (Subaru object FWHM: $1.2\arcsec$).
[\emph{See the electronic edition of the article for all ambiguous blend panels, Figures 6.1--6.341}]
}
\end{figure}

\clearpage

\section{Ellipticity of a blended pair as a function of pair separation}\label{sec:eVsSeparation}

For computational simplicity, we assume that there are two galaxies with multivariate Gaussian surface brightness profiles 
with intensities $(A_1, A_2)$, profiles defined by covariance matrices ($\Sigma_1 , \Sigma_2$), 
centroids ($\mu_1, \mu_2$), and separation 
$\theta = \left| \mu_2-\mu_1 \right|$ as illustrated in Figure~\ref{fig:galaxymodel},
the 2D flux distribution for a pair of objects is modeled as,
\begin{equation}
	F(\xv) \equiv 
	A_1 \exp \left(-\half(\xv + \muv)^T\Sigma_1^{-1}(\xv + \muv)\right)
	+
	A_2 \exp \left(-\half(\xv - \muv)^T\Sigma_2^{-1}(\xv - \muv)\right),
\end{equation}
where $\muv = \theta/2$. 

The moments are defined as,
\begin{equation}
	Q_{ij} \equiv \int\int dx_1\,dx_2\, x_{i} x_{j}\, F(\xv).
\end{equation}
For just one Gaussian term,
\begin{align}
	Q &\sim \int\int dx_1\, dx_2\, x_{i}x_{j}\exp \left(-\half(\xv-\muv)\Sigma^{-1}(\xv-\muv)\right)	
	\notag\\
	&= \int\int dy_1\, dy_2\, \left(y_i y_j + \mu_i y_j + \mu_j y_i + \mu_i \mu_j\right)
	\exp \left(-\half \yv^T \Sigma^{-1}\yv \right)
	\notag\\
	&= 2\pi \left|\Sigma\right| \left(\Sigma_{ij} + \mu_i \mu_j\right)
\end{align}
Rearranging terms,
\begin{align}
	Q_{ij} &= 2\pi \left(
	\left(A_1 \left|\Sigma_1\right| + A_2 \left|\Sigma_2\right|\right) \mu_i \mu_j 
	+ 
	\left(A_1 \left|\Sigma_1\right|\Sigma_{1, ij} + A_2 \left|\Sigma_2\right| \Sigma_{2, ij}\right)
	\right)
	\notag\\
	&\equiv M \mu_i \mu_j + C_{ij}.
\end{align}

The estimator for the ellipticity is defined from the moments,
\begin{equation}
	\left|e\right|^2 \equiv 
	\frac{ \left(Q_{11} - Q_{22}\right)^2 + Q_{12}^2}{ \left(Q_{11} + Q_{22}\right)^2}.
\end{equation}
Without loss of generality, orient the coordinate system so $\mu_2=0$. Then,
\begin{equation}
	Q_{11} = M\mu_1^2 + C_{11}, \qquad Q_{22} = C_{22}, \qquad Q_{12} = C_{12}.
\end{equation}
The ellipticity estimator becomes,
\begin{equation}
	|e|^2 = 
	\frac{ \left(M\mu_1^2 + (C_{11}-C_{22})\right)^2 + C_{12}^2}
	{ \left(M\mu_1^2 + (C_{11}+C_{22}\right)^2}.
\end{equation}
If we further assume that the galaxy ellipticities are aligned with the $x$ or $y$ axis, then $\Sigma_{12} = 0 \rightarrow C_{12}=0$, giving,
\begin{equation}\label{eq:e_est}
	|e| = \frac{M\mu_1^2 + (C_{11}-C_{22})}{M\mu_1^2 + (C_{11}+C_{22})}.
\end{equation}
When $\mu_{i} \ll \sqrt{\Sigma_{ii}}$, $|e|\sim M\mu_1^2 / (C_{11} + C_{22})$. 
In the other limit $\mu_{i} \gg \sqrt{\Sigma_{ii}}$, $|e|\rightarrow 1$.
In combination with the linear $\theta$ dependence of $dP_\mathrm{pair}$, Figure~\ref{fig:probdists} will look similar when expressed as ellipticity magnitude, except with a cubic rather than linear relation for small ellipticity.

Marginalizing over the full population of galaxy properties will smooth out this offset peak in the ellipticity distribution, however the general trend of an ellipticity distribution with larger variance for the ambiguous blended population compared to the non-blended population will remain. 
This increased variance will translate to an increase in shear noise, however it is important to note that according to our toy model this increased variance of the ambiguous blend ellipticity distribution will not lead to an additive shear systematic error since the position angle of the blended pair population will be uniformly distributed.

While we have assumed multivariate Gaussian surface brightness distributions in this derivation, the general finding that the ellipticity of the pair of objects increases with their separation  and asymptotically approaches 1 remains valid for any set of reasonable galaxy surface brightness profiles (e.g., S\'{e}rsic).

\label{lastpage}

\end{document}